# Enhanced EEG-Based Mental State Classification : A novel approach to eliminate data leakage and improve training optimization for Machine Learning

*Maxime Girard, Rémi Nahon, Enzo Tartaglione and Van-Tam Nguyen*
*LTCI, Télécom Paris, Institut Polytechnique de Paris*
maxime.girard@telecom-paris.fr

*Abstract*—In this paper, we explore prior research and introduce a new methodology for classifying mental state levels based on EEG signals utilizing machine learning (ML). Our method proposes an optimized training method by introducing a validation set and a refined standardization process to rectify data leakage shortcomings observed in preceding studies. Furthermore, we establish novel benchmark figures for various models, including random forest and deep neural networks.

*Keywords*—Attention state classification, EEG, DNN, standardization

## I. INTRODUCTION

Attention is the cognitive process involving the selection of environmental information for conscious processing [1]. Electroencephalographic (EEG) signals provide insights into the level of attention [2]. The surge in algorithms and machine learning models for attention level determination is driven by the diverse applications of attention classification. These applications span from educational tools incorporating workload monitoring to medical interventions for conditions like depression, schizophrenia, and anxiety disorders [3]. Recent studies employ signal analysis methods to leverage the frequency properties of brain signals [4][5]. Various models, such as LSTM [6], deep neural networks [4], or SVM [5][7], are trained to categorize frequencies derived from raw signals.

In this paper, we draw upon the findings of previous research, specifically [4] and [5], to present a new approach to split a public dataset containing raw EEG signals for enhanced mental state classification. We delve into the standardization method outlined in [4] and propose an alternative technique to standardize the feature vector, addressing shortcomings inherent in the prior method. Ultimately, we define updated benchmark accuracy metrics on the dataset employing various models, encompassing random forest and deep neural networks.

## II. DATASET AND METHOD

### A. Raw signals dataset used

We utilized a publicly accessible dataset proposed in [5]. This dataset comprises raw signals captured from seven electrodes of a modified EEG Epoc headset placed on the brain cavity, identified as *F3*, *F4*, *Fz*, *C3*, *C4*, *Cz*, and *Pz* in the 10-20 electrode system [8]. The sampling frequency for the records is 128Hz. The dataset was formed from a population of five individuals. For each participant, except one, seven sessions have been recorded, each lasting at least 40 minutes. For the remaining participant, only six sessions were recorded. Each recording session consisted of three phases: in the initial ten minutes, participants were asked to control a train on a simulator (*focused* state); in the next ten minutes, participants were instructed to watch the screen without controlling anything, with the instruction not to close their eyes (*unfocused* state); and after that, participants were asked to relax and close their eyes (*drowsed* state). The first two records for each participant were used for habituation and are not suitable for classification [4]. It is important to note that the dataset's size is limited, which may pose a challenge for generalization [9].

### B. Previous works

Prior studies proposed the categorization of the dataset into three classes (*focused*, *unfocused*, *drowsed*) using diverse techniques [4][5]. Within these studies, three main paradigms for training a classifier can be identified:

1. *Common-subject* paradigm. The dataset is split into two parts, independently of the subject. 80% of the data is used for training, while the remaining 20% is reserved for evaluation.

2. *Subject-specific* paradigm. The classifier is trained on a single subject. For this specific subject, 80% of the data is used for training, while the remaining 20% is reserved for evaluation.

3. *Leave-one-out* paradigm. A subject is chosen as the test subject. The data from all other subjects are employed for training, and the classifier is evaluated using the data from the designated test subject.

Research work [5] uses SVM to classify on *subject-specific* paradigm and found an accuracy of **96.70%** for subject 1. Using a 6 layers deep neural network, [4] found an accuracy of **99.9%** on the best subject. Research work [4] also found an accuracy of **99.2%** using SVM with the *common-subject* paradigm. In the *leave-one-out* paradigm, accuracy is computed by averaging the accuracy obtained when iterating the test subject across the five subjects in the dataset. In [4], the best accuracy for *leave-one-out* is **67%**.

### C. Splitting method

In this paper, our focus centers on the *leave-one-out* paradigm, which closely aligns with real-life situations, particularly given the high variability among subjects in EEG signals [10]. Unlike previous studies [4][5] where the dataset is divided into train and test sets, we introduce a new method for dataset splitting (Fig. 1.a). Here, the dataset is partitioned into three segments: train, validation, and test [15]. Similar to the *leave-one-out* paradigm in prior works [4][5], one subject is designated as the test subject. To construct a validation set, we opt to select the last record from the four remaining subjects, with the remaining records allocated for training. Recognizing the limited diversity in our dataset, we refrain from using all records of a specific subject as a validation set to ensure ample variety in the training set for generalization purposes [9]. The introduction of a validation set is essential for implementing a learning rate scheduler and an early

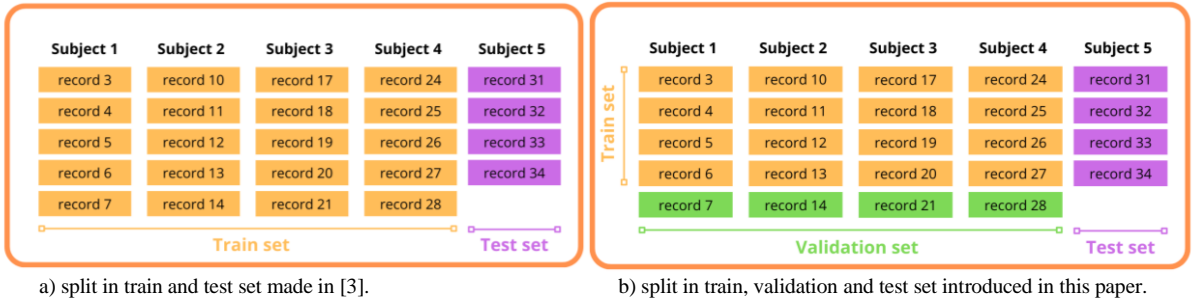

a) split in train and test set made in [3].   b) split in train, validation and test set introduced in this paper.

Fig. 1. Dataset split schemes

stopping framework, as discussed in section *III.A*. To maintain dataset balance [11], we choose to cap the records at 40 minutes, ensuring an equal time interval between drowsiness and non-drowsiness (20 minutes each).

### D. Preprocessing

We implemented the same preprocessing as detailed in [4]. For each record, we conducted a *Short-Term Fourier Transform* (*STFT*) [12]. The *window length* and *hop length* parameters were chosen within the ranges of 4 to 40 seconds and 8 to 396 samples, respectively—these parameters are further explored in section *III*. We employed the *Blackman* windowing function [13]. Following the STFT, we obtained a spectrogram where the *y*-axis signifies the frequency of the signal, spanning from 0 to 64Hz. We averaged the frequencies into 128 bins of size 0.5Hz, retaining only those within the 0 to 36Hz band, as significant changes in EEG signals manifest in this range [14]. To minimize signal noise, a 15s moving average window was applied. Ultimately, the feature vectors were constructed by concatenating the 36 features of each channel converted into *dB* at each time point. Given the 7 channels, the resulting feature vectors are of size 252.

### E. Standardization

A key focus of this paper is to examine the standardization method employed in [4] and introduce a new approach for standardizing the features. However, for the purpose of comparing the training results using the newly introduced validation set with those of [4], we maintain the same standardization <u>for the results presented in section *III*</u>. The new standardization method is proposed in section *IV*. The standardization process utilized in [4] and in section III is as follows:

1. For each record $i$ of each subject $k$, the average $\mu_{k,i}$ and standard deviation $\sigma_{k,i}$ is computed feature-wise.
2. The standardized feature vector $x_{stand}$, belonging to the record $i$ of the subject $k$, is computed doing:

$$x_{stand} = \frac{x - \mu_{k,i}}{\sigma_{k,i}} \quad (1)$$

Note that this standardization is conducted *on each record* before the dataset is split, and it is applied consistently to both the test and validation datasets.

### III. TRAINING AND RESULTS

#### A. Training method

The introduction of a validation set enables us to use a learning rate scheduler [15] and an early stopping criterion [16] based on metrics derived from evaluating the model on the validation set (*validation accuracy*, *validation loss*). It is crucial to utilize a validation set entirely distinct from the test set, as the test set should never be involved in the training process [17].

The learning rate scheduler reduces the learning rate by half when a *plateau* is reached (no improvement in validation accuracy for 3 consecutive training epochs). Employing an adaptive learning rate helps prevent oscillations and facilitates faster convergence [15]. The early stopping criterion determines when training should cease, relying on the validation accuracy metric. If no improvement is observed after 10 epochs (*patience* parameter), training is stopped. When combined with the learning rate scheduler, early stopping enables the cessation of training when the model's parameters have converged [16].

#### B. Model used

We used a neural network composed of 5 layers: the input layer, three fully connected hidden layer composed successively of 64, 128 and 64 neurons, and the output layer. This model is the same as the one called DNN4 in [4].

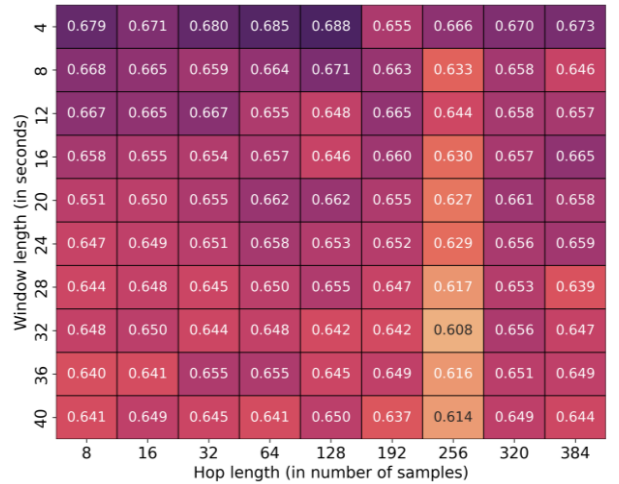

Fig. 2. Heatmap for DNN4, with (1) standardization
*x*-axis is hop length and *y*-axis is window length.

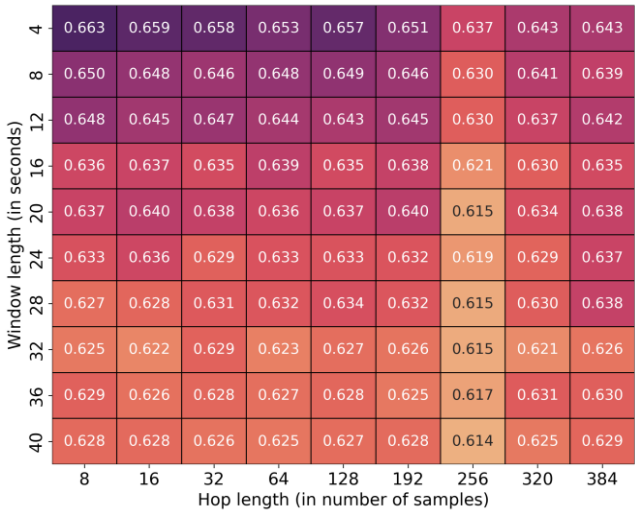

a) Random forest

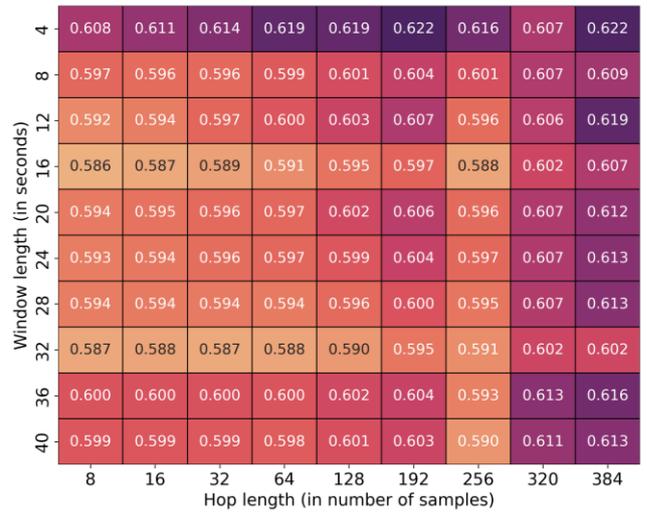

b) Support vector machine (SVM)

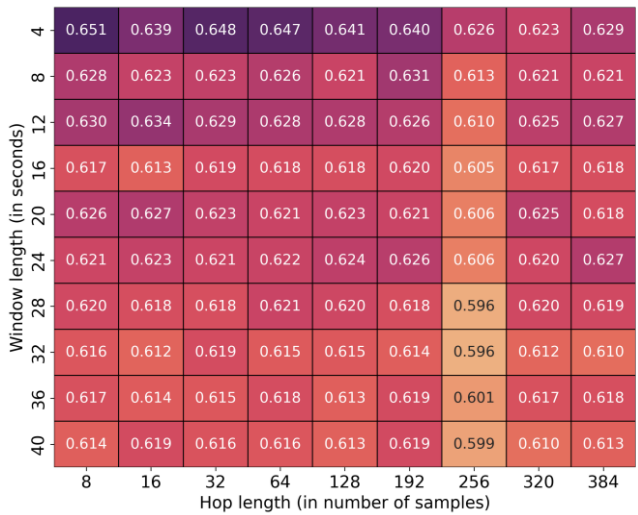

c) Extreme gradient boost (XGB)

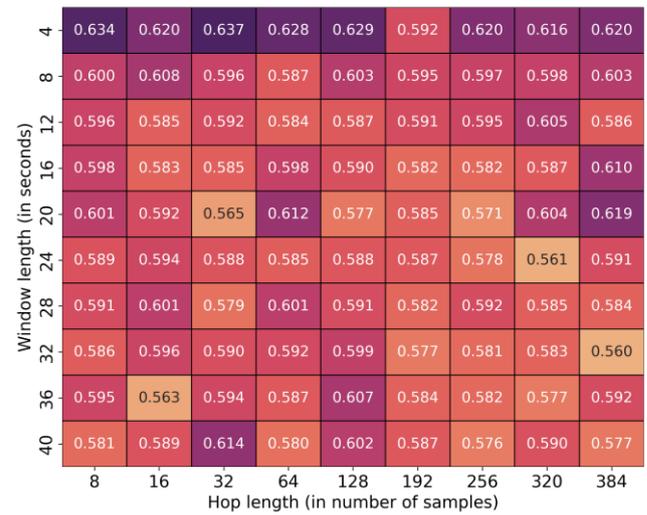

d) 4 hidden layers neural network (DNN4)

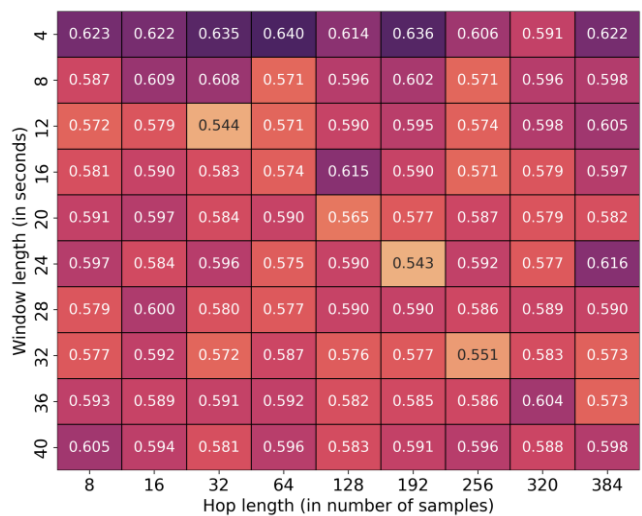

e) 6 hidden layers neural network (DNN6)

Fig. 3. Heatmaps for 5 different models –
*x*-axis is *hop length* and *y*-axis is *window length*.

## C. Results and conclusions

To determine the optimal combination of *window length* and *hop length* parameters, we generated a heatmap illustrating accuracy variations based on these two parameters (see Fig. 2). It is evident that accuracy experiences significant variations depending on these values. The most favorable results are observed in the initial rows, indicating a shorter window length. The highest accuracy achieved is **68.8%**, corresponding to a *window length* of 4 and a *hop length* of 128. Notably, for the same model, [4] reported a best accuracy of **67%**, within the same region of interest. This suggests that the incorporation of a learning rate scheduler and early stopping not only compensates for the reduction in the training set size due to the introduction of the validation set but also enables us to attain higher accuracy than before.

## IV. INTRODUCTION OF A NEW STANDARDIZATION METHOD

### A. Discussion of prior method

We detailed in section *II.E* the method used to standardize the dataset. We argue that this method is not correct and lead to data leakage [17].

Firstly, standardization should take place after the dataset is split. While it may not pose an issue in the *leave-one-out* paradigm, where splitting is based on records, it becomes a significant source of data leakage in *common-subject* and *subject-specific* scenarios, where train and test data are randomly drawn from the dataset. When computing the average and standard deviation on the entire record, including feature vectors that will belong to the test set, and applying formula (1), the feature vectors in the training set will contain data from the test set.

Furthermore, even for the *leave-one-out* paradigm, the standardization process is incorrect. The records that will be included in the test and validation sets should not be standardized individually. It is inconsistent to compute the average and standard deviation on a record belonging to the test set because, in a real-life scenario, we will not have access to future data to standardize using formula (1) [18]. More globally, we should not calculate the average and standard deviation per record but compute them for the entire training set. This ensures proper standardization for new data.

### B. Proposed standardization method

To address the data leakage issues [17] identified in the previous method discussed in section *IV.A*, we propose the following solution:

1. Before standardizing, split the dataset between train, validation, and test set.
2. Compute the average μ and standard deviation σ of the whole training set feature-wise.
3. All the feature vectors $x$, belonging either to train, test or validation set, is standardized into $x_{stand}$ by doing:

$$x_{stand} = \frac{x - \mu}{\sigma} \quad (2)$$

### C. Models used

Having resolved a data leakage issue, direct comparisons with the previously obtained results cannot be done, and we may anticipate potentially lower figures. We want to establish new reference accuracy metrics. To achieve this, we assess the accuracy of various models on our dataset. Training is conducted on random forest [19], SVM [20], XGB [21], and deep neural networks [22]. We also explore two different structures of DNN:

1. DNN4. The first DNN is composed of five layers: an input layer, three fully connected hidden layers composed of successively 512, 1024 and 512 neurons, and an output layer.
2. DNN6. The second is composed of seven layers: an input layer, five fully connected hidden layers composed of successively 512, 512, 1024, 2048 and 1024 neurons, and an output layer. Following the layer of size 2048, a dropout layer with a drop rate of 50% is incorporated to prevent overfitting [23].

The learning rate scheduler and early stopping are exclusively employed for the deep neural networks (DNNs). Nevertheless, to ensure comparability across different models, we adhere to the same split into train, test, and validation sets for all. This results in a reduced number of features for all models.

### D. Results

As in *III.C*, we trained the models for various combinations of the *window length* and *hop length* parameters. Fig. 3 presents the heatmaps for each model. The best accuracy for each model is presented in TABLE I.

TABLE I. BEST ACCURACY PER MODEL

| Model | Best accuracy | Window length | Hop length |
|---|---|---|---|
| Random Forest | 66.3% | 4 | 8 |
| SVM | 62.2% | 4 | 192 and 384 |
| XGB | 65.1% | 4 | 8 |
| DNN4 | 63.7% | 4 | 32 |
| DNN6 | 64.0% | 4 | 64 |

The highest accuracy is achieved by the random forest model, attaining **66.3%.** We must note that neural networks demand more data than random forest to reach a comparable level of accuracy [24]. Therefore, the notably limited size of our dataset might elucidate why random forest outperforms the other models, particularly neural networks, for our specific task.

We achieve the highest accuracy of **64.0%** for DNN6, a performance not far behind the random forest. This accuracy could likely be enhanced with a larger training set. The optimal region for *window length* and *hop length* parameters seems to be in the first row and first columns of Fig. 3.e. Subsequent research efforts should focus on exploring this area of interest.

## V. CONCLUSION

This paper introduced a new method for splitting a public dataset of EEG signals used in mental state classification, incorporating a validation set to improve upon previous findings reported in [4]. Furthermore, we addressed and corrected the standardization method presented in this prior work, establishing new benchmark metrics for various models, including the top-performing random forest and promising neural networks. The obtained results are highly encouraging, indicating the potential for further exploration of preprocessing techniques for attention level classification in EEG signals. For example, investigating the application of wavelets may lead to enhanced accuracy.